# Origin of the Non-Linear Pressure Effects in Perovskite Manganites: Buckling of Mn-O-Mn Bonds and Jahn-Teller Distortion of the $MnO_6$ Octahedra Induced by Pressure

Z. Chen[1], T. A. Tyson[1], K. H. Ahn[1], Z. Zhong[2], J. Hu[3]

[1]*Department of Physics, New Jersey Institute of Technology, Newark, NJ 07102*

[2]*National Synchrotron Light Source, Brookhaven National Laboratory, Upton, NY, 11973*

[3]*X17C of NSLS, Cars, University of Chicago, Upton, NY 11973*

High-pressure resistivity and x-ray diffraction measurements were conducted on $La_{0.85}MnO_{3-\delta}$ to ~6 GPa and ~7 GPa, respectively. At low pressures the metal-insulator transition temperature ($T_{MI}$) increases linearly up to a critical pressure, P* ~ 3.4 GPa, followed by reduction of $T_{MI}$ at higher pressure. Analysis of the bond distances and bond angles reveal that a bandwidth increase drives the increase of $T_{MI}$ below P*. The reduction of $T_{MI}$ at higher pressures is found to result from Jahn-Teller distortions of the $MnO_6$ octahedra. The role of anharmonic interatomic potentials is discussed.

The $RE_{1-x}A_xMnO_3$ (RE= rare earth and A = Ca, Sr, Ba) mixed valence ($Mn^{3+}$ ($d^4, t_{2g}^3 e_g^1$)/ $Mn^{4+}$ ($d^3, t_{2g}^3 e_g^0$)) perovskite system exhibits complex and intriguing properties and an understanding of the basic physics of these materials has still not been realized [1,2]. This system is under detailed study both from a fundamental science and as well as application perspectives. It has been found that a strong coupling exists among the lattice, spin, and electronic degree of freedom that is manifested by complex phase diagrams. The properties of manganites depend strongly on subtle changes in the structure and chemistry of the system induced by changing the RE or A site ion size [3]. Separating the changes in structure from the changes in valence can be accomplished by using a series of RE cations of varying size at fixed RE/A ratio. However, precise control of the stable crystalline form produced by substitution is not typically possible. Modification of the structure by cation substitution may alter the system in unpredictable ways.

A controllable way to explore the effect of strain or pressure on these systems is to apply hydrostatic pressure and then to measure the transport and structural properties [4]. Not many high-pressure measurements have been conducted on manganites. Early temperature dependent studies of high magnetoresistance phase for pressures were conducted below 2 GPa and predicted a linear increase of the metal-insulator (MI) transition temperature with pressure [5,6,7]. More recent studies on the changes in the metal-insulator transition temperature ($T_{MI}$) at pressures up to 6 GPa reveal that an optimal pressure is reached beyond which the transition temperature decreases with pressure [8,9,10]. Meneghini's *et al* [11] conducted studies (up to 15 GPa) on the effects of high pressures on the transport properties of $La_{1-x}Ca_xMnO_3$ (x=1/4), but the electrical transport data were presented only between ~190K and ~380K – limiting the access to the MI transition at high pressures.

The self-doped system $La_{0.85}MnO_{3-\delta}$ shares similar transport, magnetic and structural characteristics with the chemically doped system $La_{1-x}Ca_xMnO_3$ in the ferromagnetic phase [12], such as the classic metal-insulator transition from a high-temperature paramagnetic insulating phase to a low-temperature ferromagnetic metallic phase, yet it is chemically simple. It achieves a high magnetization and has a low resistivity at low temperature. In order to understand the origin of the changes in the electronic structure of manganites at both high and low pressures and to compare this system with the classical ion doped systems, we conducted detailed high-pressure transport and structural studies.

A polycrystalline sample of $La_{0.85}MnO_{3-\delta}$ was synthesized in air by the conventional solid-state reaction as in Ref. [12] with three calcinations cycles. The cation ratio La/Mn ~0.85 was determined by the Ionic Coupled Plasma method. High-pressure synchrotron x-ray powder diffraction measurements were conducted and Rietveld refinements were carried out to extract the detailed atomic structure following the approaches in Ref. [8].

The magnetization at ambient-pressure at 4 K (0.5 Tesla) yielded a saturation value of $3.5\mu_B$/Mn site (close to the theoretical limit of $3.56\mu_B$/Mn) with a Curie temperature (244.0 ± 1.0K) near the metal-insulator transition temperature (248.0 ± 1.0K) as shown in Fig. 1(c).

Figure 1(a) shows the electrical resistivity versus temperature for pressures ranging from ambient to 5.8 GPa. At ~3.4 GPa (P*, the critical pressure), $T_{MI}$ reaches its maximum value of ~280 K. In Fig. 1(b), we show the resistivity at 300K and 90 K as a function of pressure. Note the increase in resistivity at high pressure above P*. Similar results have been found for $La_{0.60}Y_{0.07}Ca_{0.33}MnO_3$ [8], $Nd_{1-x}Sr_xMnO_3$ (x=0.45, 0.5) [9], $Pr_{1-x}Ca_xMnO_3$ (0.25<x<0.35) [10] and $La_{0.67}Ca_{0.33}MnO_3$ [13]. In this Letter, we present the non-linear structural changes induced by pressure in the self-doped system, their relations to transport properties, and the physical basis for a saturation of physical properties at P* in the full class of manganites.

Below P*, the transition temperature $T_{MI}$ increases linearly with the pressures at the rate of 10.1±0.4K/GPa and the resistivity near room temperature decreases with the pressures. This result agrees with most of the earlier investigations of pressure dependent resistivity for mixed valance perovskite samples, such as $La_{1-x}Ca_xMnO_3$ (0.2<x<0.4) [14,15] for pressures below 2GPa. Above ~3.4 GPa, $T_{MI}$ is found to decrease sharply with increasing pressure at the approximate rate of -74.0±8.0K/GPa. (The pressure dependence of $T_{MI}$ is given in Fig. 4(d)).

Figure 2 displays representative x-ray diffraction patterns over the pressure range ambient to ~ 7 GPa at room temperature. Higher pressure structural measurements revealed that no change in space group is observed up to 11 GPa. At ambient-pressure, the lattice parameters for space group I2/a are a= 7.8012 (13) Å, b= 5.5265 (1) Å, c= 5.4818 (1) Å, β= 90.828(7) °, which compare well with Ref. [12].

In Fig. 3, we show the pressure dependence of the lattice parameters and the unit-cell volume. Note the changes of slope in the lattice parameters near ~3.4 GPa. The unit-cell volume is fit by using the first order Birch-Murnaghan equation of state over the entire pressure range, which is shown as the solid line in Fig. 3(b). Note that significant deviation of the points from the line particularly also near P*. This suggests that a single equation of state does not adequately describe the pressure dependence.

By Rietveld refinements of the XRD data, the pressure dependence of the structure was obtained. Figure 4 shows the pressure dependence of the average bond distance of $<d_{Mn-O}>$ (Fig. 4(a)), the bond angle of $<\theta_{Mn-O2-Mn}>$ in the b-c plane (Fig. 4(b)), the $MnO_6$ octahedral distortion $\delta = \sqrt{\frac{1}{6}\sum(d_{Mn-O} - <d_{Mn-O}>)^2}$ (Fig. 4(c)), $T_{MI}$ (Fig. 4(d)), the charge-carrier bandwidth $W$ (Fig.

4(e)), and the activation energy $E_a$ (taken from fits to the resistivity above $T_{MI}$) (Fig. 4(f)) based on $\rho = \rho_0 T \exp(E_a/kT)$ [16].

In the observed pressure range, the average tilt angles of $MnO_6$ octahedra do not change significantly. However, the average Mn-O bond distance contracts monotonically below the critical pressure P* and maintains a constant value above P*. In the framework of the double-exchange model, the magnetic and transport properties are determined by the charge-carrier bandwidth *W*, which depends on the Mn-O bond distance and the Mn-O-Mn bond angle through the overlap integral between Mn 3d orbital via O 2p orbital. In our work, we have used an approximate expression, $W \propto \cos(1/2(\pi - <\theta_{Mn-O-Mn}>))/d_{Mn-O}^{3.5}$, based on the radial [17] and the angular dependence (Mn-O-Mn) of the overlap matrix [18, 19], as shown in Fig. 4(e). Note that the bandwidth is increasing below P*, however, decreases very slowly above P*. The $MnO_6$ octahedral distortions were found to be constant at the low-pressure range but suddenly increase at higher pressures (Fig. 4(c)).

These results show that, below P*, the increase in $T_{MI}$ with pressure is due to the increase in the electron bandwidth, which reaches its maximum at P*. Moreover increases in pressure yield enhance local distortions, which trap the $e_g$ conduction electrons and produce an insulating phase. This can be further seen from the enhancement of the resistivity for pressure above P* in Fig. 1(c). In fact, the increase of the resistivity seen at low temperature is reminiscent of percolation, which occurs through the introduction of an insulating phase into a metallic phase [20]. The activation energy $E_a$ for hopping (in the high-temperature region) exhibits a minimum at P*, which is consistent with the maximum of the bandwidth.

To explain our data below *P**, we consider a typical inter-atomic potential, *V(r)*, between the RE and the oxygen (O) ion in perovskite manganites, for example, a potential energy curve similar to the Lennard-Jones potential with a strong repulsive core and a weakly attractive tail. In $RE_{1-x}A_xMnO_3$ perovskites when the tolerance factor ($t = \frac{(<R_A> + R_O)}{\sqrt{2}(R_{Mn} + R_O)}$ and $R_i$ are the atomic radii) is less than 1 [19], the distance between RE/A and O before Mn-O-Mn buckling, $r_0$, would be larger than the distance with the minimum potential energy. Buckling of a Mn-O-Mn bond moves an O ion closer to the RE on one side and farther apart from the RE on the other side, which gives rise to two different RE-O distances. The buckling would occur only if the net energy decreases upon buckling, in other words, $d^2V(r)/dr^2|_{r=r0} < 0$. If the RE-O potential is purely harmonic, the buckling of Mn-O-Mn bonds would not occur. It is the anharmonicity of the potential energy that reduces the total potential energy upon buckling. The effect of external pressure is to move the point $r_0$ closer to the harmonic region of the potential (more precisely, to the inflection point of the potential), which reduces the energy gain associated with buckling and, therefore, buckling itself as observed below $P^*$ in our data. Such straightening of the Mn-O-Mn bond angles as well as the reduced volume under pressure contribute to the rapid decrease of the average Mn-O distance, as observed for pressure below $P^*$. Both effects associated with the Mn-O-Mn bond angle and the Mn-O distance add up to increase the effective hopping amplitude of the $e_g$ electrons on Mn sites, which explains the increase of $W$ and $T_{MI}$ below $P^*$.

We now examine possible explanations for our data at pressures above $P^*$. Once the Mn-O distance is reduced into the region of a hard-core potential, the Mn-O distance is very hard to compress any further by external pressure. We believe that $P^*$ marks such critical pressure. Above $P^*$, the average Mn-O distance remains almost constant as shown in our data. The only way to reduce the Mn-Mn distance and, therefore, the volume of the unit cell by applied pressure is to buckle Mn-O-Mn bonds, which explains the observed increase of the buckling (i.e. reduced Mn-O-Mn angle) above $P^*$. Our data also shows the rapid increase of the Jahn-Teller distortion above $P^*$. The Jahn-Teller coupling can be considered in terms of hybridization between Mn $e_g$ orbitals and oxygen *p* orbitals, which increases for a shorter average Mn-O distance. Therefore, the Jahn-Teller coupling would be enhanced for smaller average Mn-O distances. The appearance of the strong Jahn-Teller distortion only above $P^*$ not below $P^*$, in spite of similar Mn-O-Mn buckling, demonstrates the importance of the average Mn-O distance on the strength of the Jahn-Teller coupling. The rapid decrease of $T_{MI}$ above $P^*$ is consistent with the reduced electron mobility due to the increased buckling and the Jahn-Teller distortion.

In summary, high-pressure resistivity and x-ray diffraction measurements have been performed on $La_{0.85}MnO_{3-\delta}$. We find that the metal-insulator transition temperature ($T_{MI}$) increases with pressure up to a critical value, P* ~3.4 GPa, beyond which $T_{MI}$ decreases with increasing pressure. Analysis of the Mn-O bond distances and Mn-O-Mn bond angles reveal a close correlation between structural distortions and transport properties. In particular, the bandwidth increase drives the increase of $T_{MI}$ for pressure below P*. The reduction of $T_{MI}$ at higher pressures is found to result from the Jahn-Teller distortions of the $MnO_6$ octahedra and the localization of 3d electrons. The general trend is expected to be a characteristic feature of manganites. We find the anharmonic interatomic potential plays an important role for the structural changes in Mn-O distances, Mn-O-Mn bond angles, and the Jahn-Teller distortions under pressure.

We are indebted to Prof. J. B. Goodenough (University of Texas at Austin) for important contributions to the discussions on the pressure dependence of the bandwidth. The high-pressure x-ray diffraction measurements were conducted at Brookhaven National Laboratory's NSLS beamline X17B1. This research was funded by NSF DMR-0512196.

**Figure captions**

**Fig. 1.** (Color online) (a) Temperature dependent resistivity for pressures varying from ambient to 5.8 GPa. Open symbols show the low-pressure regime with increasing $T_{MI}$ while the closed symbols show the high-pressure regime. The arrows indicate the transition temperatures (inflection point). (b) Resistivity at 300 K and 90K as a function of pressure. (c) Ambient-pressure resistivity ($\Omega$*cm) and magnetization ($\mu_B$/Mn) at a magnetic field of 0.5T as a function of temperature for $La_{0.85}MnO_{3-\delta}$.

**Fig. 2**. (Color online) (a) The synchrotron x-ray powder diffraction patterns at various pressures up to 6.9 GPa, at room temperature, with a vertical dotted line as a guide to the eyes. (b) Diffraction pattern (crosses) at the 4.0 GPa with its Rietveld fit (solid) and residual (lower line).

**Fig. 3.** (a) Lattice parameters of $La_{0.85}MnO_{3-\delta}$ at the ambient temperature as a function of pressure. (b) Unit-cell volume versus pressure up to 7 GPa with a curve depicting the first order Birch-Murnaghan fit for equation of state.

**Fig. 4**. Pressure dependence of structural and electronic parameters. The average Mn-O distance (a), average Mn-O2-Mn bond angle (b), the $MnO_6$ octahedral distortion (c), the metal-insulator transition temperature $T_{MI}$ (d), the electron bandwidth W with arbitrary units, (e) and activation energy for charge-carrier hopping $E_a$ above $T_{MI}$ (f), are plotted as a function of pressure.

**Figure 1, Z. Chen *et al*., Physics Review Letter.**

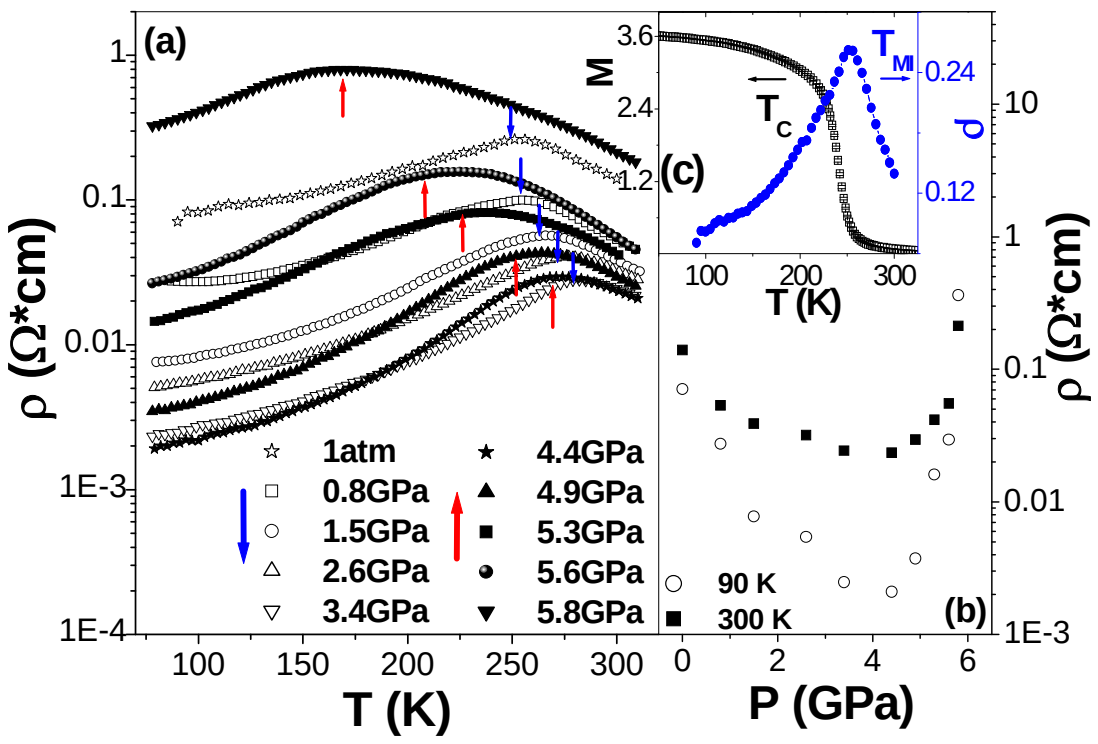

**Figure2, Z. Chen *et al*., Physics Review Letter.**

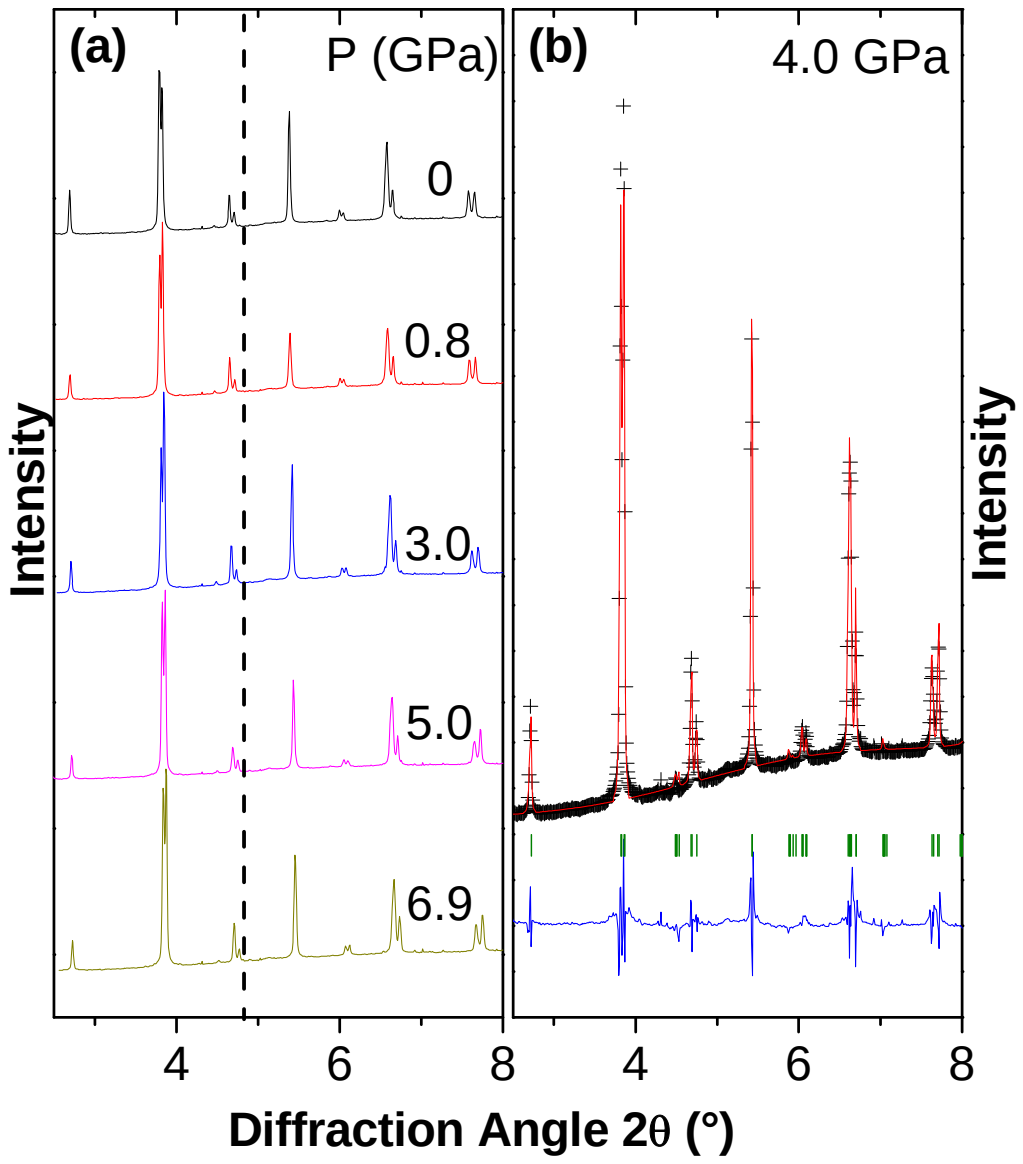

**Figure 3, Z. Chen *et al*., Physics Review Letter.**

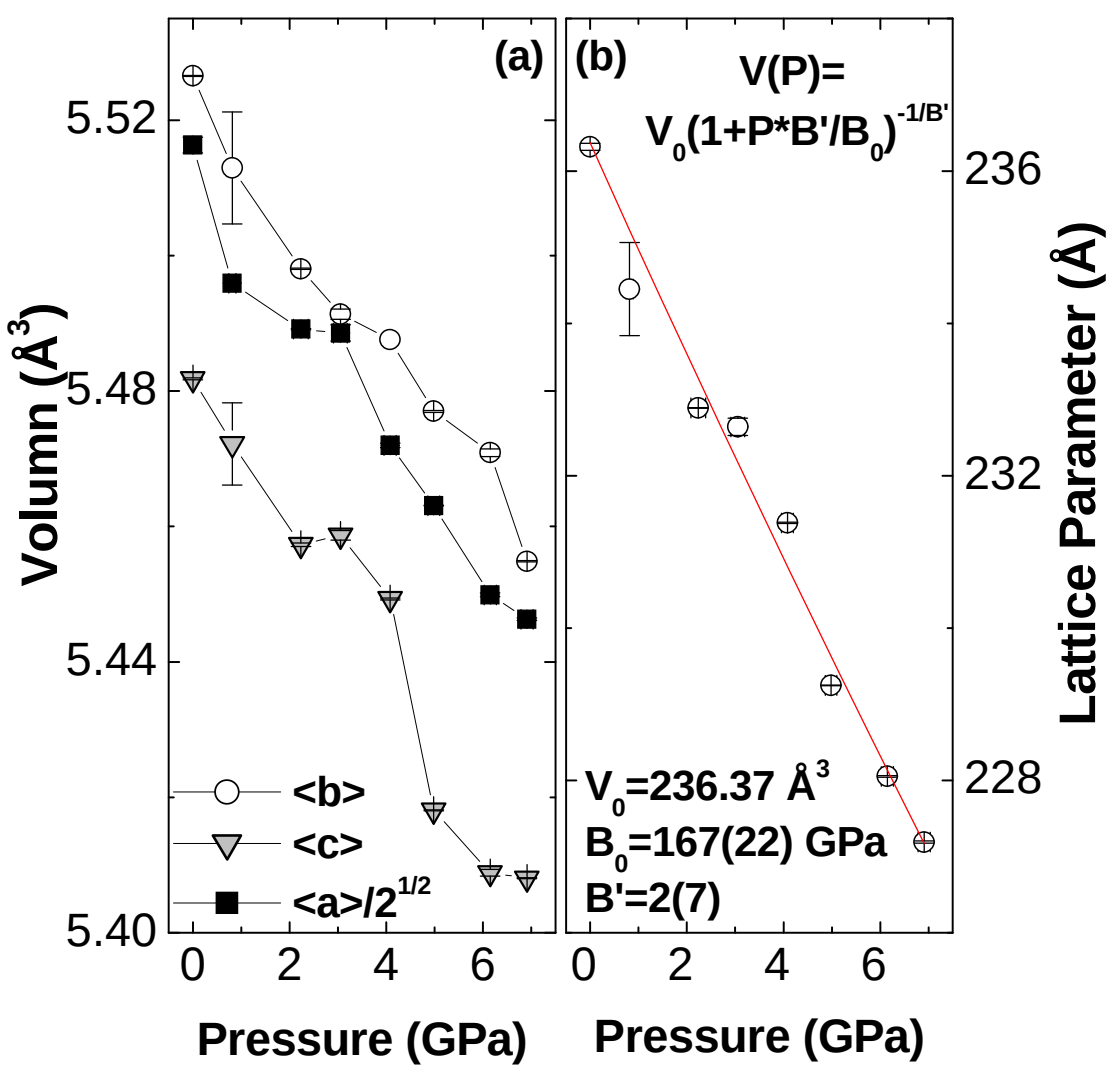

**Figure 4, Z. Chen *et al*., Physics Review Letter.**

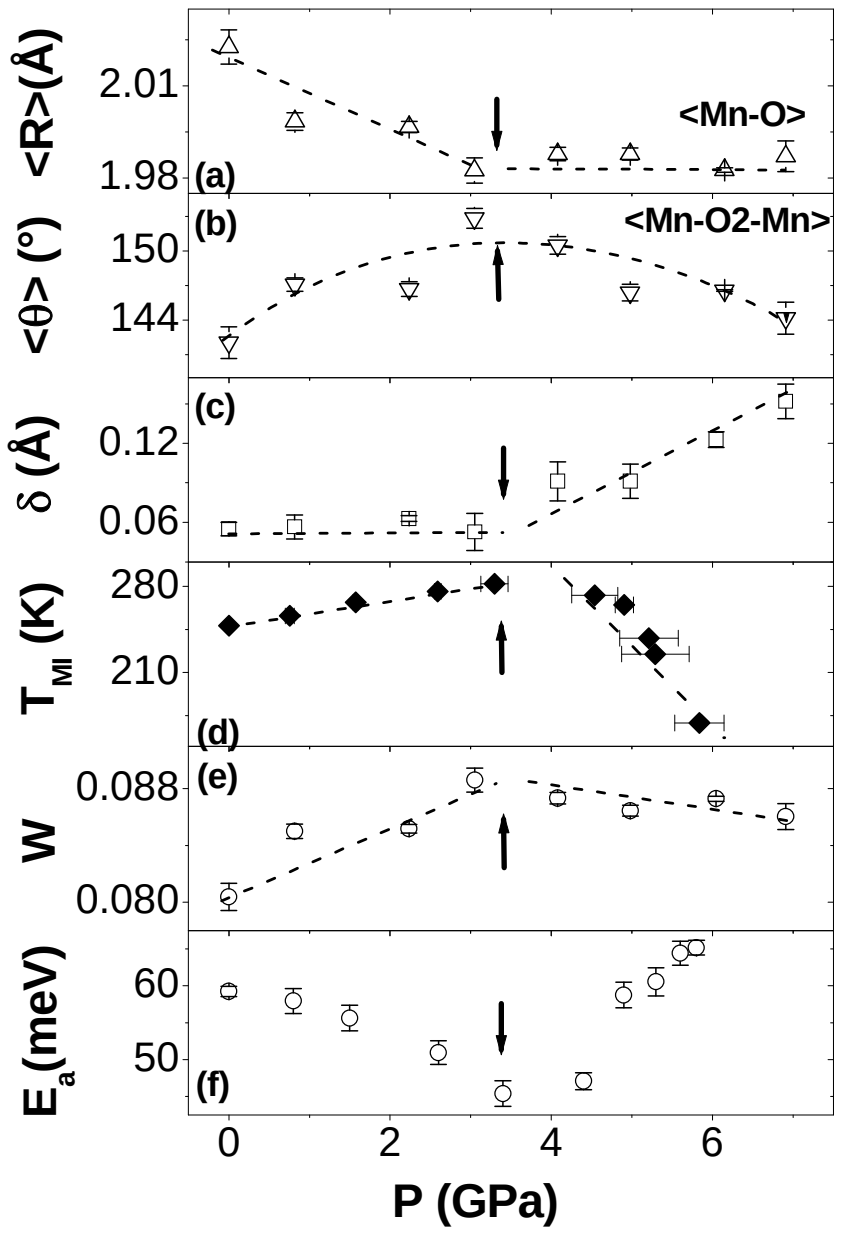